\documentclass[aps,pre,reprint,superscriptaddress,amsmath,showpacs,amssymb]{revtex4-1}
\usepackage{amssymb}
\usepackage{tipa}
\usepackage{bm,epsfig,mathrsfs,amsmath,amssymb,graphicx}
\usepackage{epsfig,amsmath,graphicx,amssymb}
\usepackage{graphicx}
\usepackage{dcolumn}
\usepackage{bm}
\usepackage{color}
\usepackage[toc,title,titletoc,header]{appendix}

\begin{document}

\newcommand\uleip{\affiliation{Institut f\"ur Theoretische Physik, Universit\"at Leipzig,  Postfach 100 920, D-04009 Leipzig, Germany}}
\newcommand\chau{\affiliation{ 
 Charles University,  
 Faculty of Mathematics and Physics, 
 Department of Macromolecular Physics, 
 V Hole{\v s}ovi{\v c}k{\' a}ch 2, 
 CZ-180~00~Praha, Czech Republic 
}}

\title{Maximum efficiency of low-dissipation refrigerators at arbitrary cooling power}
\author{Viktor Holubec}\email{viktor.holubec@mff.cuni.cz}\uleip\chau
\author{Zhuolin Ye}\email{zhuolinye@foxmail.com}\uleip

\begin{abstract}
We analytically derive maximum efficiency at given cooling power for Carnot-type low-dissipation refrigerators. The corresponding optimal cycle duration depends on a single parameter, which is a specific combination of irreversibility parameters and bath temperatures. For a slight decrease in power with respect to its maximum value, the maximum efficiency exhibits an infinitely-fast nonlinear increase, which is standard in heat engines, only for a limited range of parameters. Otherwise, it increases only linearly with the slope given by ratio of irreversibility parameters. This behavior can be traced to the fact that maximum power is attained for vanishing duration of the hot isotherm. Due to the lengthiness of the full solution for the maximum efficiency, we discuss and demonstrate these results using simple approximations valid for parameters yielding the two different qualitative behaviors. We also discuss relation of our findings to those obtained for minimally nonlinear irreversible refrigerators.
\end{abstract}

\maketitle
\date{\today}

\section{Introduction}

The laws of energy conservation and non-decrease of entropy of the universe, cornerstones of classical thermodynamics developed during 19th century, imply universal upper bounds on efficiencies of thermodynamic machines such as heat engines, heat pumps, and refrigerators~\cite{Callen2006}. They are reached by idealized machines operating under reversible conditions, with vanishing net entropy production. The advantage of these results is their generality. The disadvantages are omnipresent dissipation looses in real machines, rendering their reversible operation difficult, and even more importantly, the fact that reversible conditions correspond to practically negligible output power~\cite{Holubec2017}. 

These issues triggered less general, but more practical branch of research based on various models of irreversible and/or finite-time thermodynamics, which is efficiency of thermodynamic machines at maximum power. Starting with the works on performance of nuclear power plants by Yvon, Chambadal, and Novikov \cite{Yvon1955,Chambadal1957,Novikov1958} later popularized by Curzon and Ahlborn~\cite{curzon1975efficiency}, this model-based research attracted a considerable attention during last fifty years, and is still lively today. Efficiency at maximum power has been studied for endoreversible~\cite{curzon1975efficiency, rubin1982optimal, chen1989unified}, low-dissipation \cite{ME-low-dissipation, gonzalez2020energetic, wang2012coefficient}, linear irreversible~\cite{VandenBroeck2005,benenti2011thermodynamic, izumida2014work}, minimally nonlinear irreversible~\cite{Izu2015, Izu-EPL-heat, Izu-EPL-refri},
quantum~\cite{uzdin2014universal, abah2012single, rossnagel2014nanoscale}, and Brownian~\cite{schmiedl2007efficiency, segal2008stochastic, jarzynski1999feynman} models.

During recent years, based on the above models, yet another, even more practice-oriented, branch of research started, optimisation of efficiency at given power. For vanishing power the maximum efficiency equals to the reversible limit, and for maximum power to efficiency at maximum power. 
Below, we address this task, previously solved for various heat engines~\cite{Holubec2017} but only minimally nonlinear irreversible refrigerators \cite{zhang2018coefficient, Long2018}, for Carnot-type low-dissipation refrigerators.

In next two Secs.~\ref{sec:model} and \ref{sec:td}, we introduce in detail the considered model and define variables describing its thermodynamic performance. In Sec.~\ref{sec:EMP}, we review the corresponding result on efficiency at maximum power. Our main results on maximum efficiency at given power are given in Sec.~\ref{sec:MAXCOP}. We conclude in Sec.~\ref{sec:conclusion}. The relation between the low-dissipation and minimally nonlinear irreversible models is discussed in Appendix~Sec.~\ref{appx:MIM}.
 
\section{Model and assumptions}
\label{sec:model}

\begin{figure}[htbp]
\centering
\includegraphics[width=0.65\columnwidth]{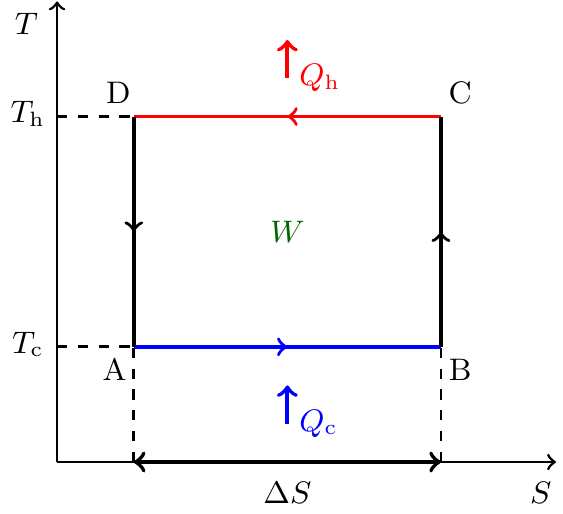}
\caption{Thermodynamic $T$-$S$ (bath temperature-system entropy) diagram of the considered Carnot refrigeration cycle. The fridge uses the input work $W$ to extract heat $Q_{\rm c}$ from the cold bath at temperature $T_{\rm c}$ during the cold isotherm (AB, blue). The used work and the extracted heat are then dumped as heat $Q_{\rm h} = Q_{\rm c} + W$ into the hot bath at temperature $T_{\rm h}$ during the hot isotherm (CD, red). The input work equals the enclosed area, $Q_{\rm c} = T_{\rm c} \Delta S$, and $Q_{\rm h} = T_{\rm h} \Delta S$ only if the cycle is performed reversibly. Otherwise, the work is larger and the extracted heat smaller leading to a decreased efficiency (coefficient of performance) of the machine. The branches BC and DA (black) of the cycle are adiabats.}
\label{fig:T-S}
\end{figure}

We consider a refrigerator operating along a finite-time Carnot cycle of duration $t_{\rm p}$ depicted and described in detail in Fig.~\ref{fig:T-S}. We assume that in the limit of infinitely slow driving, $t_{\rm p} \to \infty$, the fridge operates reversibly and its finite-time performance is captured by the so-called low-dissipation (LD) assumption~\cite{ME-low-dissipation}
\begin {equation}
Q_{\rm h} = T_{\rm h}\Delta S + \frac{\sigma _{\rm h}}{t_{\rm h}},
\label{qh}
\end {equation}
\begin {equation}
Q_{\rm c} = T_{\rm c}\Delta S - \frac{\sigma _{\rm c}}{t_{\rm c}},
\label{qc}
\end {equation}
for total amounts of heat interchanged with the individual reservoirs during the cycle. The ratio $\sigma _{\rm h}/(t_{\rm h}T_{\rm h})$  measures an excess in the total amount of entropy $Q_{\rm h}/T_{\rm h} - \Delta S$ produced during the hot isotherm due to its finite duration $t_{\rm h}$, and similarly for $\sigma _{\rm c}/(t_{\rm c}T_{\rm c})$. We assume that the adiabatic branches interconnecting the isotherms are ideal and thus the net amount of entropy $\Delta {S_{\rm tot}}$ produced per cycle is solely given by the dissipation due to the heat transferred to the two reservoirs during the isotherms:
\begin{equation}
 \Delta {S_{\rm tot}} = \frac{Q_{\rm h}}{T_{\rm h}} - \frac{Q_{\rm c}}{T_{\rm c}} =\frac{\sigma _{\rm h}}{t_{\rm h}T_{\rm h}} + \frac{\sigma _{\rm c}}{t_{\rm c}T_{\rm c}}.\label{delta-entropy-total}
\end{equation}
The fridge hence operates reversibly if the isotherms are infinitely slow (and thus $t_{\rm p} \to \infty$) or if the so-called irreversibility parameters $\sigma _{\rm h}$ and $\sigma _{\rm c}$ vanish.

Interestingly, this simple model, where all thermodynamically important details about the system dynamics are described by the irreversible parameters, represents quite well two general realistic setups, justifying the considerable attention it received in recent literature~\cite{ME-low-dissipation, gonzalez2020energetic, wang2012coefficient, holubec2016maximum,holubec2015efficiency,Holubec2015Erratum,Gonzalez-Ayala2018,Gonzalez-Ayala2019,gonzalez2020energetic}. First, Eqs.~\eqref{qh} and \eqref{qc} can be interpreted as formal expansions of the interchanged heats in the inverse cycle duration $1/t_{\rm p}$. Therefore, they should be generally valid for slowly, but not quasi-statically, driven systems. Indeed, the decay of total dissipated heat with the inverse of duration was theoretically predicted for various quantum and classical setups \cite{Sekimoto1997,Zulkowski2015,Cavina2017,Holubec2020} and observed in various experiments \cite{mart15,Ma2019}. The second situation, where the assumption \eqref{qh} and \eqref{qc} holds for arbitrary cycle duration, are overdamped Brownian systems driven by special time-dependent protocols (usually minimizing dissipated energy during the isotherms \cite{schmiedl2007efficiency,Holubec2014,holubec2015efficiency,muratore2015efficient,Long2018}). While a similar optimisation might also be performed for other systems, we are not aware of such results.

Furthermore, models of thermal machines utilizing the LD assumption can \emph{exactly} be mapped to the minimally nonlinear irreversible (MNI) model operating under the tight coupling condition~\cite{Izu-EPL-heat, Izu-EPL-refri, Izu2015, iyyappan2018general}. This broadly used model of irreversible thermodynamics generalizes the standard linear irreversible model~\cite{VandenBroeck2005,Ryabov2016} by including terms describing dissipation of the input work due to an internal friction (or, in case of thermoelectric machines, resistivity \cite{apertet2013efficiency}), which are proportional to the irreversibility parameters. Even though this model can describe also cyclically operating systems~\cite{Proesmans2016}, it does not incorporate any obvious periodicity and thus it is usually interpreted as operating in a non-equilibrium steady-state. On the other hand, the LD model naturally describes machines operating cyclically. Therefore, thermal machines described by the two models are usually optimised differently. The natural control parameter for MNI models is the external force $X_1$, corresponding to the (scaled) input work $W/T_{\rm h}$ in the LD model, or, equivalently, the flux $J_1$ conjugated to $X_1$, which stands in the mapping to the LD model for the inverse cycle-duration $1/t_{\rm p}$. Since the LD models are not only optimised with respect to $t_{\rm p}$ but also with respect to distribution of this total duration among the individual branches of the cycle, the obtained optimal performance in the two models usually differ. The notable exception are bounds on performance obtained by further optimising with respect to the irreversible parameters. Then the two optimisation procedures coincide and the results obtained within the two models agree. For more details, see Appendix~\ref{appx:MIM}.

\section{Power and efficiency}
\label{sec:td}

Central quantities describing performance of a refrigerator are its cooling power, $P$, and efficiency, $\varepsilon$, often referred to as the coefficient of performance (COP). The cooling power is defined as heat extracted from the cold bath per cycle over the cycle duration,
\begin {equation}
P =\frac{Q_{\rm c}}{t_{\rm p}}=\frac{T_{\rm c}\Delta S}{t_{\rm p}}-\frac{\sigma _{\rm c}}{t_{\rm c}t_{\rm p}}, 
\label{power}
\end {equation}
where we have applied the LD assumption~\eqref{qc}. The COP measures cost of the cooling in units of input work, $W = Q_{\rm h} - Q_{\rm c}$, used to pump the heat from the cold bath, 
\begin {equation}
\varepsilon=\frac{Q_{\rm c}}{W}=\frac{\varepsilon _{\rm C}}{1+T_{\rm h}\varepsilon _{\rm C} {\Delta S_{\rm tot}}/{(P t_{\rm p})}}.
\label{eta}
\end {equation}
First glance at these definitions reminds us the textbook knowledge that simultaneous optimisation of power and COP is not possible (textbooks usually deal with heat engines, but the situation with refrigerators is the same). Maximum COP, $\varepsilon_{\rm C} = T_{\rm c}/(T_{\rm h}-T_{\rm c})$, is attained under reversible conditions ($\Delta S_{\rm tot} = 0$) when the term $T_{\rm h}\varepsilon _{\rm C} \Delta S_{\rm tot}/(P t_{\rm p})$ in the denominator of Eq.~\eqref{eta} vanishes. And, even though recent theoretical results on thermodynamics of small systems allowing unprecedented control of the intrinsic relaxation times show that power corresponding to $\varepsilon_{\rm C}$ can even diverge~\cite{Campisi2016,Polettini2017,Pietzonka2018,Holubec2018}, it is doomed to be negligible compared to its maximal value~\cite{Holubec2017}. 

In practice, we thus always have to resort to a compromise between power and COP. To this end, various ad hoc trade-off figures of merit of refrigerators have been proposed.
Examples are the $\chi$ criterion \cite{de2012optimal, yan1990class, yuan2014coefficient}, the $\Omega$ criterion \cite{hernandez2001unified, de2013low, long2014performance}, and the ecological criterion \cite{angulo1991ecological, yan1996optimization, ust2007performance}.
However, none of these tell us what we really want to know: what is the exact cost of running a refrigerator with a specific cooling power, which is usually fixed by our needs (for example size of the space that should be cooled). The optimisation task of practical interest is thus to find maximum COP for a given cooling power, i.e. to show under which conditions is this cooling power cheapest. With respect to heat engines, this task already gained considerable attention in the literature~\cite{Holubec2017}. Expressions for maximum efficiency at given power were derived for quantum thermoelectric heat engines~\cite{Whitney2014,Whitney2015}, LD heat engines~\cite{holubec2016maximum, See-AppendixC}, MNI heat engines~\cite{Long2016}, a stochastic heat engine based on an uderdamped harmonic oscillator \cite{Dechant2017}, and using general linear response theory \cite{Ryabov2016}. With respect to refrigerators, the treasury of results for general models is not so overflowing, with a notable exception of results for MNI refrigerators~\cite{Long2018}. 

Below, we derive maximum COP at given power for the LD model defined above. Our bounds~\eqref{eta-duration2} on the maximum COP agrees with those obtained by Long et al.~\cite{Long2018} for MNI refrigerators. This is because, in these limiting cases, the two, generally different, optimisation procedures agree. The rest of our results differ from those for MNI refrigerators quantitatively, but the most interesting qualitative features of the obtained maximum COP are preserved. And thus our discussion below might interest also readers of Ref.~\cite{Long2018}.

\section{COP at maximum cooling power}
\label{sec:EMP}

The values of cooling power accessible to a refrigerator are bounded by 0 and the maximum power, $P^\star$. A natural starting point for calculating maximum COP at fixed power is thus determination of $P^\star$ for LD refrigerators, which was done in Ref.~\cite{hernandez2015time}. Since peculiarities of the derivation strongly affect qualitative behavior of maximum COP at fixed power, we review it in detail. 

We aim to maximize the cooling power~\eqref{power} as function of the cycle duration $t_{\rm p}$ and its division among the individual branches. To this end, we assume, without loss of generality, that the sum of durations of the adiabatic branches is proportional to the total duration of the isotherms, $t_{\rm i} = t_{\rm h} + t_{\rm c}$, so that $t_{\rm p} = a t_{\rm i}$ with $a \ge 1$. This assumption allows us to simplify the calculations and it can easily be relaxed. Maximum power is obviously obtained for $a=a^\star=1$ (adiabats infinitely faster than isotherms). Since the parameter $a$ does not influence the COP~\eqref{eta}, we keep it at this value for the rest of our discussion. Even though such infinitely fast adiabatic branches seem strange at first glance, they were realized in experiments with Brownian heat engines~\cite{Blickle2012}. Together with infinitely fast adiabats might come an issue with bringing the system far from equilibrium, thus effectively breaking the regime of validity of the LD model. However, this can be avoided by properly adjusting the value of the control parameter (for example volume or stiffness of a potential) and temperature at the ends of the adiabatic branches~\cite{Sekimoto2000,Holubec2014}. Readers who nevertheless feel uncomfortable with setting $a=1$ can redefine the power for the rest of the paper as $a P$.
Furthermore, we introduce the dimensionless parameter 
\begin{equation}
 \alpha = t_{\rm h}/t_{\rm i}  \in [0,1]
 \label{eq:alpha}
\end{equation}
measuring relative duration of the hot isotherm.

Maximizing the cooling power in Eq. (\ref{power}) with respect to $t_{\rm i}$ gives~\cite{hernandez2015time}
\begin{eqnarray}
t _{\rm i,\alpha}^ *  &=& \frac{{2{\sigma _{\rm c}}}}{{{\left(1-\alpha\right) T_{\rm c}}\Delta S}},
\label{taup*}\\
P_{{\alpha}}^ *  &=& \frac{{{{\left(1-\alpha\right)\left({{T_{\rm c}}\Delta S} \right)}^2}}}{{4{\sigma _{\rm c}}}}, 
\label{Ptp}\\
\varepsilon _{\alpha}^ *  &=&
\frac{\varepsilon _{\rm C}}{ 2+\varepsilon _{\rm C}+\sigma \varepsilon _{\rm C}\left(1/\alpha-1\right)},
\label{etatp}
\end{eqnarray}
where we have introduced the irreversibility ratio
\begin{equation}
\sigma\equiv\sigma_{\rm h}/\sigma_{\rm c}.
\label{eq:dissipaton_ratio}
\end{equation}
With decreasing $\alpha$, the partially optimised cooling power~\eqref{Ptp} monotonously interpolates between 0 (attained for $\alpha = 1$, $t _{\rm i,\alpha}^ * = \infty$ and $\varepsilon_{\alpha}^ * = \varepsilon_{\rm C}/(2+\varepsilon_{\rm C})$, note that this process is not reversible even-though the cycle duration diverges) and its maximum, reached for $\alpha=\alpha^*=0$~\cite{hernandez2015time}. The resulting maximum power and the corresponding duration of the isothermal branches thus read
\begin{eqnarray}
P^* &=& \frac{\left(T_{\rm c}\Delta S\right)^2}{4\sigma_{\rm c}},
\label{final-power}\\
t _{\rm i}^ *  &=& \frac{{2{\sigma _{\rm c}}}}{{{ T_{\rm c}}\Delta S}}.
\label{final-taup}
\end{eqnarray}
With increasing irreversibility parameter $\sigma_{\rm c}$,
the maximum power and $1/t_{\rm i}$ monotonously interpolates between 0 and $\infty$. In contrast, the COP at maximum power, $\varepsilon^ *$, reads
\begin{eqnarray}
\varepsilon_-^ * &=& 0 \quad \quad \quad \,\,\, {\rm for} \quad \sigma>0
\label{final-COP-}\\
\varepsilon_+^ * &=& \frac{\varepsilon_{ \rm C}}{2+\varepsilon_{\rm C}} \quad {\rm for} \quad \sigma=0 
\label{final-COP+}
\end{eqnarray}
and thus it exhibits a discontinuity at $\sigma = 0$~\cite{hernandez2015time}, \textcolor{black}{which should be understood in the sense that $\sigma_{\rm h} \ll \sigma_{\rm c}$.} 
This discontinuity is caused by the requirement $\alpha^* = 0$, \textcolor{black}{ which should be understood in the sense that the duration of the hot isotherm is negligible compared to that of the cold one, i.e. $t_{\rm h} \ll t_{\rm c}$.} Then the total entropy production~\eqref{delta-entropy-total} diverges unless the irreversibility parameter $\sigma$ is set to zero before $\alpha$.

\textcolor{black}{Actually, if one does not set} $\alpha = \alpha^* = 0$ exactly in the derivation, but instead consider a limiting process $\alpha \to \alpha^*$, he can get efficiencies at maximum power, $\epsilon^*$, within the bounds $[\varepsilon_-^ *,\varepsilon_+^ *]$. For example, assuming that $\lim_{\alpha\to\alpha^*}\sigma/\alpha = k$ and thus $\sigma=k\alpha$, $k>0$, the efficiency at maximum power reads 
\begin{equation}
\epsilon^* = \frac{\varepsilon _{\rm C}}{ 2+(1+k)\varepsilon _{\rm C}}.
\end{equation}

 Since the Eqs.~\eqref{final-COP-} and \eqref{final-COP+} do not depend on $\sigma$, they represent lower and upper bounds on COP at maximum power of LD refrigerators. For MNI refrigerators, $\varepsilon_\pm^ *$ describe bounds on COP at maximum power~\cite{Izu2015,Long2018}, obtained as extreme values of~\eqref{etatp} as function of $\sigma$. Thus the discontinuity found in the LD model, caused by optimisation with respect to $\alpha$, is  not present in the MNI model.

  In closing this section, we should discuss how reasonable is taking the limit $\sigma \to 0$, leading to the nontrivial value $\varepsilon_+^ *$ of COP at maximum power, from a physical perspective. However, we postpone this discussion to Sec.~\ref{sec:ultimateCOP} and continue with optimisation of COP at fixed power.

\section{Maximum COP at arbitrary cooling power}
\label{sec:MAXCOP}

From technical reasons~\cite{holubec2015efficiency,holubec2016maximum,Ryabov2016,Holubec2017}, it is advantageous to study the maximum COP at fixed power using the dimensionless loss in power (with respect to the maximum power),
\begin{equation}
\delta P \equiv \frac{P-P^*}{P^*} \in [-1,0],
\label{eq:dp}
\end{equation}
and the dimensionless duration (of the isotherms),
\begin{equation}
\tau= \frac{t_{\rm i}-t_{\rm i}^*}{t_{\rm i}^*} \in [-1,\infty].
\label{eq:dtau}
\end{equation}
The loss in power vanishes for $P = P^\star$ and assumes its minimum value $-1$ if the power $P$ is negligible compared to $P^\star$~\cite{Holubec2017}. \textcolor{black}{The definition~\eqref{eq:dp} physically means that we measure energy flows in units of maximum power~\eqref{final-power} and thus we effectively fix value of $\sigma_{\rm c}$.} The duration $\tau$ equals to $-1$ for $t_{\rm i} = 0$ and it is negative (positive) for $t_{\rm i} < t_{\rm i}^*$ ($t_{\rm i} > t_{\rm i}^*$). Since we are interested in maximum COP at fixed power and longer cycles in general allow for larger COPs, our intuition suggests (and the calculation below proofs) that we can focus on positive values of $\tau$ only.

Fixing the cooling power (or, equivalently, $\delta P$) creates a dependence between the duration, $\tau$, and relative duration of the hot isotherm, $\alpha$. Using Eq.~(\ref{power}), we find that
\begin{equation}
\alpha=1+\frac{1}{\left(1+\delta P\right)\tau^2+2\delta P\tau+\delta P-1}.
\label{alphadeltap}
\end{equation}
Using further the definition~\eqref{eq:alpha}, implying that $0 \le \alpha \le 1$, we find that the above formula makes sense only for a limited interval of $\tau$:
\begin{equation}
-\frac{\sqrt{-\delta P}}{{1+\sqrt { -\delta P}}} \le \tau  \le \frac{\sqrt{-\delta P}}{{1-\sqrt { -\delta P}}}.
\label{tau-duration}
\end{equation}
The COP~\eqref{eta} in these new variables reads
\begin{equation}
\varepsilon  = \frac{\tau ^3 +A_{1,3}\tau ^2 +A_{0,3}\tau  + A_{0,1}}{- {\tau ^3} + A_{1/{\varepsilon_+^ *},-3} \tau^2 + B_{3,4,1}\tau + B_{1,2,-1}},
\label{cop-deltap-equation}
\end{equation}
with $A_{k,l} = (k +l \delta P)/(1 + \delta P)$ and $B_{k,l, m}= [-k{\left( \delta P \right)}^2+ \left( l/\varepsilon _{\rm C} + 1+\sigma \right)\delta P+ m \sigma]/ \left(1 + \delta P \right)^2$, and we will now find its maximum as function of $\tau$.

\subsection{Bounds}
\label{sec:ultimateCOP}

For fixed $\tau$, $\delta P$, and $\varepsilon_{\rm C}$, the COP~\eqref{cop-deltap-equation} is a monotonously decreasing function of $\sigma$. Analytically, this follows by noticing that $\partial \varepsilon/ \partial \sigma < 0$. Intuitively, it can be understood as follows. \textcolor{black}{As noted above, $\sigma_{\rm c}$ is fixed by the chosen energy unit $P^*$ and thus $\sigma$ is solely determined by $\sigma_{\rm h}$.} COP~\eqref{eta} monotonously decreases with increasing entropy production $\Delta S_{\rm tot}$, which is, for fixed dissipation during the cold isotherm, monotonously increasing function of $\sigma_{\rm h}$.

The lower bound on COP is thus obtained in the limit of infinitely irreversible hot isotherm ($\sigma = \infty$). Then, $\Delta S_{\rm tot}/P$ in Eq.~\eqref{eta} diverges and the maximum attainable COP vanishes, regardless values of $\tau$ and $\delta P$. Fortunately, the upper bound on COP, obtained if the hot isotherm is reversible ($\sigma = 0$), is positive. In this case, Eq.~\eqref{cop-deltap-equation} can be simplified to
\begin{equation}
\varepsilon = \left[\frac{2\left(1 + \varepsilon_{\rm C}\right)}{\left(1+\tau\right)\varepsilon_{\rm C}\left(1+\delta P\right)}-1\right]^{-1}.\label{eta-tau-sigma-0}   
\end{equation}
For the allowed values~\eqref{tau-duration} of $\tau$, this function monotonously increases and thus the upper bound on COP is obtained by setting $\tau = \sqrt{-\delta P}/(1-\sqrt { -\delta P})$. In agreement with the result derived for MNI refrigerators~\cite{Long2018}, we find that the maximum COP at fixed power, $\varepsilon^{\rm opt} = \varepsilon^{\rm opt}(\delta P)$, is bounded as
\begin{equation}
0 \le \varepsilon^{\rm opt} \le \frac{\varepsilon_{\rm C}\left(1+\sqrt{-\delta P}\right)}{2+\varepsilon_{\rm C}\left(1-\sqrt{-\delta P}\right)} \equiv \varepsilon^{\rm opt}_+.
\label{eta-duration2}
\end{equation}
All known bounds on maximum efficiency at fixed power for heat engines~\cite{holubec2016maximum,Ryabov2016,Holubec2017,Whitney2014,Whitney2015,Dechant2017,Long2016} exhibit an infinite gain in efficiency (with respect to the efficiency at maximum power) when the engines operate at powers infinitely smaller than $P^*$, in symbols $\partial \eta^{\rm opt}/\partial \delta P|_{\delta P = 0} \to \infty$. The upper bound $\varepsilon^{\rm opt}_+$ on $\varepsilon^{\rm opt}$ in Eq.~\eqref{eta-duration2} shows qualitatively the same large gain in COP. The corresponding relative gain in COP for small $\delta P$ reads
\begin{equation}
\frac{\varepsilon^{\rm opt}_+ - \varepsilon^*_+}{\varepsilon^*_+} = \left(1 + \varepsilon^*_+ \right)\sqrt{-\delta P} + \mathcal{O}\left(\delta P\right),
\label{eq:rel_gain_0}
\end{equation}
where $\mathcal{O}\left(\delta P\right)$ denotes a correction of order $\delta P$. Thus the derivative of the relative gain with respect to $\delta P$ diverges with $\delta P \to 0-$ as $1/\sqrt{-\delta P}$. This is a general behavior expected for a COP near maximum power if the later is determined by vanishing derivative with respect to a control parameter $x$~\cite{holubec2016maximum,Ryabov2016,Dechant2017}. Indeed, if $\partial P/\partial x|_{P=P^*} = 0$ (in the present setting, $x$ stands for $\alpha$ or $\tau$) one would expect that expansions of power and efficiency around the maximum power $P^*$ read $\delta P \approx - x^2/c^2$ and $\varepsilon - \varepsilon^* = |a| x$, leading to the relation $\varepsilon - \varepsilon^* = |a c| \sqrt{-\delta P }$. In the present case, however, the maximum power~\eqref{final-power} does not correspond to a a vanishing derivative with respect to $\alpha$. As a result, the described ``universal'' behavior can be for LD refrigerators observed for small parameters $\sigma$ and $\varepsilon_{\rm C}$ only, as suggested by behavior of bounds~\eqref{eta-duration2} and discussed in the following two sections. 

In closing this section, let us review how (physically) reasonable are the limiting values 0 and $\infty$ of the irreversibility ratio, leading to the bounds~\eqref{eta-duration2}. To this end, there is a handful of microscopic models yielding reasonable expressions for $\sigma$. For relatively broad class of slowly driven systems (described by generalized Markovian master equation with a symmetric protocol for hot and cold isotherms), the irreversibility ratio assumes the form $\sigma = \left(T_{\rm h}/T_{\rm c}\right)^{1-\xi}$, where $\xi$ stands for the exponent in the bath spectral density~\cite{Cavina2017}. The limit $\sigma\to 0$ thus corresponds to an infinitely super-Ohmic bath ($\xi\to \infty$), while the opposite limit $\sigma\to \infty$ is obtained for an infinitely sub-Ohmic bath ($\xi\to -\infty$). Obviously, neither of such strongly-diverging spectral densities (and thus also the corresponding values of $\sigma$) make much physical sense. For overdamped Brownian dynamics with time-dependent driving optimized to minimize the dissipated work, the irreversibility ratio is given by the ratio $\sigma = \mu_{\rm c}/\mu_{\rm h}$ of mobilities~\cite{schmiedl2007efficiency}. Since the infinite mobility is not compatible with assumptions of overdamped dynamics~\cite{schmiedl2007efficiency, holubec2015efficiency,Holubec2015Erratum,Holubec2017,Holubec2018}, meaningful possibilities to reach the limiting values of $\sigma$ are the vanishing mobility $\mu_{\rm c}$ during the cold isotherm ($\sigma=\infty$) or vanishing mobility during the hot isotherm ($\sigma = 0$). Such conditions can indeed be realized. One ensuing technical problem is that with decreasing mobility increases relaxation time of the system, and thus one has to resort to a stronger driving to get the same performance~\cite{Holubec2018}. To conclude, realising the limiting values of the irreversibility ratio exactly in the lab is practically impossible, but these regimes can theoretically be reasonably approximated. Nevertheless, this can be quite expensive and thus, for real practical applications, it is important to study behavior of $\varepsilon^{\rm opt}$ also for non-extreme values of the irreversibility ratio, which is a topic of the next section.   

\subsection{Arbitrary parameters}
\label{sec:exact_bound}

\begin{figure}[htbp]
\includegraphics[width=0.95\columnwidth]{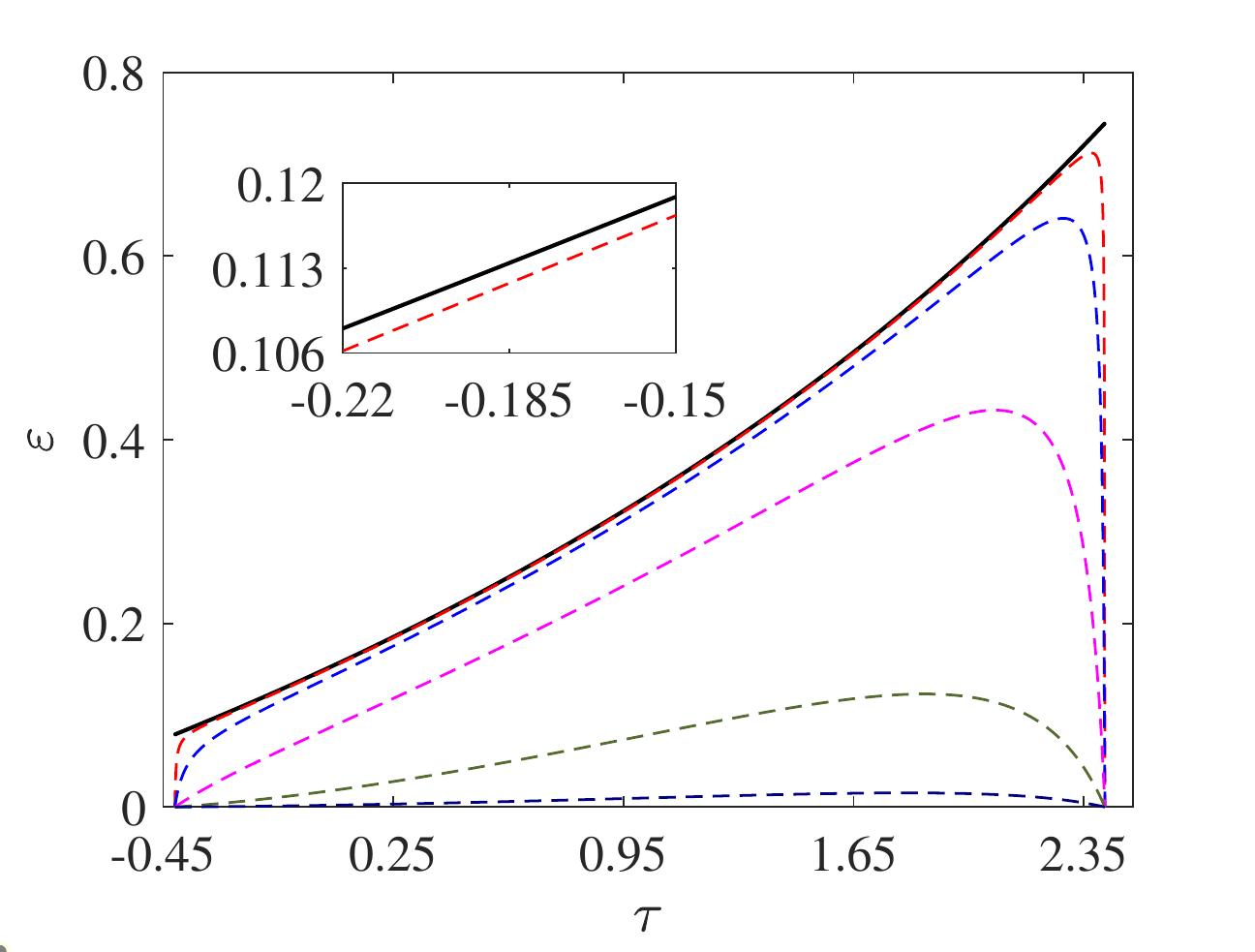}
\caption{COP~\eqref{cop-deltap-equation} as function of $\tau$ for six values 0, 0.01, 0.1, 1, 10, and 100 of $\sigma$ increasing from the top black solid line to the lowermost broken line. The inset shows that the maximum COP is attained at $\sigma = 0$. Parameters taken: $\delta P=-0.5$ and $\varepsilon_{\rm C}=1$.}
\label{FixedDP}
\end{figure}

In Fig.~\ref{FixedDP}, we show the COP \eqref{cop-deltap-equation} as function of the duration $\tau$ for six values of the irreversibility ratio. For $\sigma = 0$, $\varepsilon$ indeed monotonously increases. For all larger $\sigma$, it develops a peak at a position $\tau^{\rm opt} < \sqrt{-\delta P}/(1-\sqrt{-\delta P})$, which can be determined from the condition $\partial \varepsilon /\partial \tau|_{\tau=\tau^{\rm opt}}  = 0$. Explicitly, it reads
\begin{multline}
(\tau^{\rm opt}) ^4 + \tilde{A} (\tau^{\rm opt})  ^3 + \tilde{B}_{6+3\tilde{\sigma},2+2\tilde{\sigma},-\tilde{\sigma}} (\tau^{\rm opt})  ^2\\ 
+ \tilde{B}_{4+3\tilde{\sigma},-2\tilde{\sigma},-\tilde{\sigma}}\tau^{\rm opt}  
+ \tilde{B}_{1+\tilde{\sigma},-2\tilde{\sigma},0} = 0,
\label{max-cop-derive}
\end{multline}
where the the coefficients $\tilde{A} = [(4+\tilde{\sigma})\delta P + \tilde{\sigma}]/(1+\delta P)$ and $\tilde{B}_{k,l,m} = (k\delta P^2 + l \delta P + m)/(1+\delta P)^2$ depend on $\sigma$ and $\varepsilon _{\rm C}$ only through the combination $\tilde \sigma = \sigma /\left({\frac{1}{{{\varepsilon _{\rm C}}}} + 1}\right)$. For given loss in power, the optimal duration is thus solely determined by $\tilde \sigma$, which monotonously increases both with $\sigma$ and $\varepsilon _{\rm C}$. 

\begin{figure}[htbp]
\includegraphics[width=0.9\columnwidth]{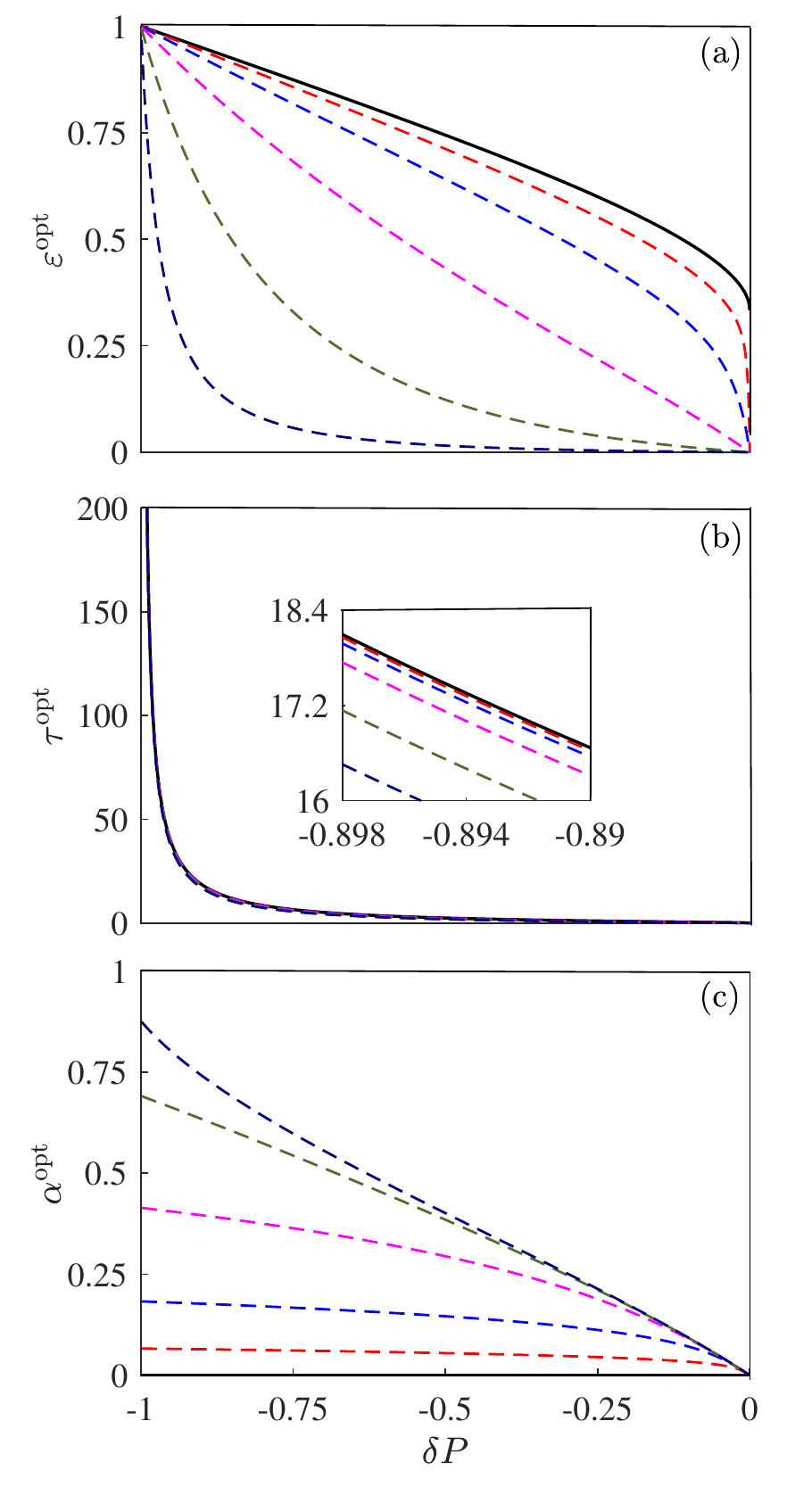}
\caption{(a) The optimal COP $\varepsilon^{\rm opt}$ as a function of $\delta P$ for six values
0, 0.01, 0.1, 1, 10, and 100 of $\sigma$ increasing from the top black solid line to the lowermost broken line. Panels (b) and (c) show the corresponding parameters $\tau^{\rm opt}$ and $\alpha^{\rm opt}$. The inset in panel (b) magnifies the differences between the individual curves. Note the inverse ordering of the curves in the panel (c). We took $\varepsilon _{\rm C} = 1$.} \label{Exact}
\end{figure}

The quartic equation (\ref{max-cop-derive}) has four roots and can be analytically solved using the Ferrari's method \cite{cardano19681545}. The physical optimal duration $\tau^{\rm opt}$ is given by the root in the interval~\eqref{tau-duration}, which can be determined by inserting some specific values of $\delta P$ and $\tilde \sigma$ into the formal expressions for the roots. Even though the ensuing expression is far too long and cumbersome to be more enlightening than a numerical solution, it can be used as a basis of various approximations explaining the qualitative behavior of $\tau^{\rm opt}$ (or $\alpha^{\rm opt}$) and $\varepsilon^{\rm opt}$, depicted in Fig.~\ref{Exact}.

Specifically, the maximum COP, shown in Fig.~\ref{Exact} (a), exhibits a sharp increase with power near the maximum power only for small values of $\sigma$. In agreement with the discussion in the preceding section, the rate of this increase $-\partial \varepsilon^{\rm opt}/ \partial \delta P |_{\delta P = 0}$ actually decreases with $\sigma$ from $\infty$ (for $\sigma \to \infty$) to 0 (for $\sigma = 0$). For large values of $\sigma$, the maximum COP exhibits a fast increase (similar to that of $\varepsilon^{\rm opt}$ near $P = P^*$ for small $\sigma$) close to the vanishing power, where the COP attains its ultimate upper bound $\varepsilon_{\rm C}$. 
While the described dependence of $\varepsilon^{\rm opt}$ on $\sigma$ is significant, the optimal duration $\tau^{\rm opt}$ in Fig.~\ref{Exact} (b) changes with $\sigma$ only slightly, always monotonously interpolating between 0 for $\delta P = 0$ and $\infty$ for $\delta P = -1$. This suggests that a reasonable approximation of $\tau^{\rm opt}$ substituted for $\tau$ in Eq.~\eqref{cop-deltap-equation} might lead to an excellent approximation of $\varepsilon^{\rm opt}$. \textcolor{black}{The optimal relative duration of the hot isotherm shown in Fig.~\ref{Exact} (c) is fixed by $\tau$ and $\delta P$  through Eq.~\eqref{alphadeltap}. Contrary to that of $\tau^{\rm opt}$, the dependence of $\alpha^{\rm opt}$ on $\sigma$ is significant. Let us note that  a similar situation occurs also for LD heat engines~\cite{holubec2016maximum}.}
To get a more analytical and quantitative grasp of the described qualitative behavior of the maximum COP, we now derive several approximate formulas valid in the two regions of $\sigma$ described above. 

\subsection{Approximations}

\subsubsection{Small irreversibility ratio}

Expanding the  exact optimal duration $\tau^{\rm opt}$ and COP $\varepsilon^{\rm opt}$, obtained using Eq.~\eqref{max-cop-derive}, up to the first order in $\tilde \sigma$, we find that, up to a correction $\mathcal{O}\left(\tilde{\sigma}\right)$,
\begin{eqnarray}
\tau^{\rm opt} &\approx&
\frac{\sqrt{-\delta P}}{1-\sqrt{-\delta P}}-\frac{\sqrt{\tilde\sigma}}{2 (-\delta P)^{1/4}},
\label{eq:tausigmas}\\
\varepsilon^{\rm opt} &\approx& \varepsilon^{\rm opt}_+ -\frac{2 (1+\varepsilon_{\rm C}) (\varepsilon^{\rm opt}_+)^2  (1-\sqrt{-\delta P}) \sqrt{\tilde \sigma}}{\varepsilon_{\rm C} (-\delta P)^{1/4}  (1+\sqrt{-\delta P})}.
\label{eq:COPsigmas}
\end{eqnarray}
The expansion~~\eqref{eq:COPsigmas} explodes to $-\infty$ for $\delta P\to 0$ and thus it makes sense for reasonably large $-\delta P$ only. Similar divergence is present for all other terms in the series. This, mathematically undesirable, sharp decrease of the correction term for small $-\delta P$ describes the jump in the COP at maximum power~\eqref{final-COP-}--\eqref{final-COP+} from $\varepsilon_+^*$ for $\sigma = 0$  to $0$ for $\sigma >0$. Note that the approximate optimal duration~\eqref{eq:tausigmas} exhibits a similar behavior.

\subsubsection{Large temperature difference}

The above approximation is valid for small $\tilde{\sigma}$ attained both for large temperature differences (small $\varepsilon_{\rm C}$) and small irreversibility ratios $\sigma$. For small $\varepsilon_{\rm C}$ the expression~\eqref{eq:COPsigmas} can be further simplified to
\begin{equation}
\frac{\varepsilon^{\rm opt}}{\varepsilon_{\rm C}} =\frac{\left(1+\sqrt{-\delta P}\right)}{2}-\frac{\left(1+\delta P\right){\tilde \sigma}^{1/2}}{2\left(-\delta P\right)^{1/4}}+\mathcal{O}\left(\varepsilon_{\rm C}\right).\label{app-smallcop}
\end{equation}
Interestingly, the first term above is the same as that in LD~\cite{holubec2016maximum}, linear irreversible~\cite{Ryabov2016}, and MNI~\cite{Long2016} heat engines. 

\subsubsection{Large irreversibility ratio}

\begin{figure*}[htbp]
\includegraphics[trim={1cm 0cm 4cm 0},width=0.95\linewidth]{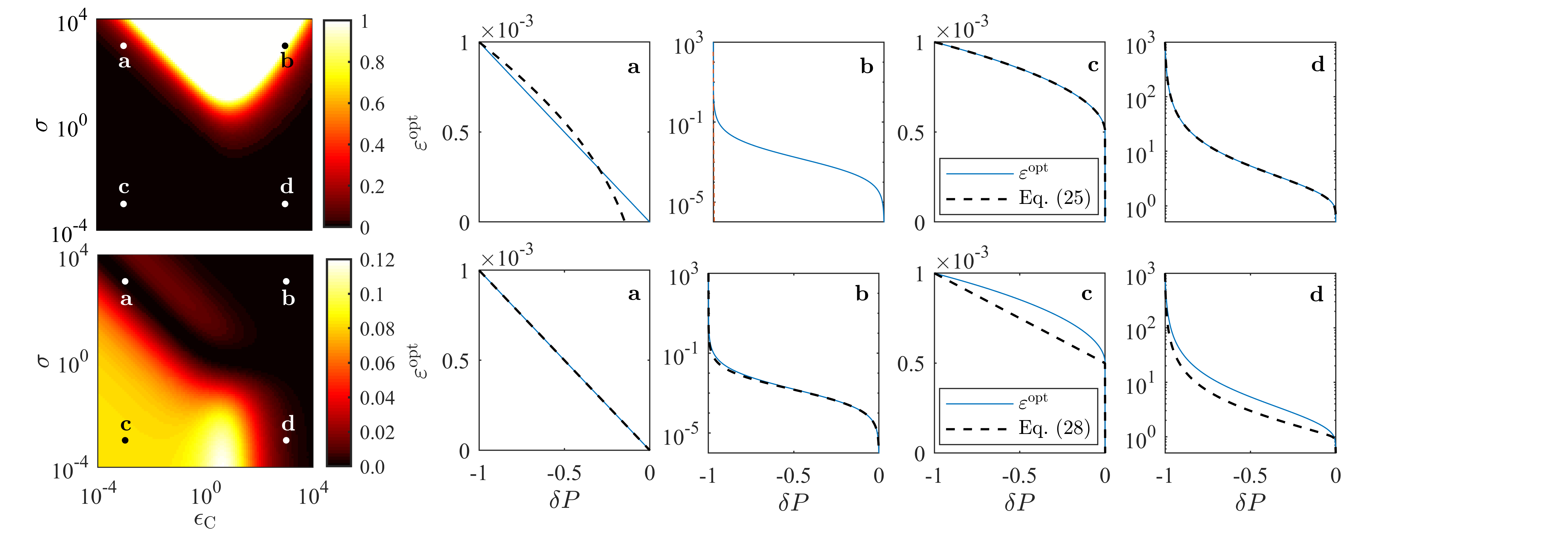}
\caption{Performance of the approximations \eqref{eq:COPsigmas} (top) and \eqref{app-largecop} (bottom) of the exact maximum COP. The left panels depict the error~\eqref{eq:Error} (color code) as function of $\varepsilon_{\rm C}$ and $\sigma$. The remaining panels show the individual approximate functions (broken lines) and the exact $\varepsilon^{\rm opt}$ (solid lines) for the points {\bf a}, {\bf b}, {\bf c}, and {\bf d}, depicted in the leftmost figures. Their coordinates are {\bf a} $= (10^{-3}, 10^{3})$, {\bf b} $= (10^{3}, 10^{3})$, {\bf c} $= (10^{-3}, 10^{-3})$, and {\bf d} $= (10^{3}, 10^{-3})$. For the parameters {\bf b}, Eq.~\eqref{eq:COPsigmas} yields positive values only near the left boundary of the corresponding figure. For parameters {\bf c} and {\bf d} above and {\bf a} and {\bf b} below, the curves almost perfectly overlap.} 
\label{Test}
\end{figure*}

Let us now turn to the case of large irreversibility ratio. Up to the leading order in $\tilde{\sigma}$, solution to Eq.~\eqref{max-cop-derive} reads 
\begin{equation}
\tau^{\rm opt}=-\frac{2\delta P}{1+\delta P}.
\label{eq:tau_opt_etaCsmall}
\end{equation}
Interestingly, the same expression is obtained for $\tilde{\sigma} = 1$, and thus, for example, for an infinitely small temperature difference ($\varepsilon_{\rm C} \to \infty$) and $\sigma = 1$.
Substituting this $\tau^{\rm opt}$ for $\tau$ in Eq. (\ref{cop-deltap-equation}) leads to the expression for maximum COP at fixed power
\begin{equation}
\varepsilon^{\rm opt}\approx\frac{\delta P \left(1-\delta P\right)\varepsilon_{ \rm C}}{2\delta P+\left(1+\delta P\right)\left(\delta P-\sigma\right)\varepsilon_{\rm C}},
\label{app-largecop}
\end{equation}
which is exact for $\tilde{\sigma} = 1$ and $\infty$, and which can be expected to give a good approximation of $\varepsilon^{\rm opt}$ for all $\tilde{\sigma} \in [1,\infty]$. The expansion of Eq.~\eqref{app-largecop} up to the first order in $\delta P$ reads
\begin{equation}
\varepsilon^{\rm opt}\approx    -\frac{\delta P}{\sigma}+\mathcal{O}\left(\delta P\right).
\label{eq:larges_approx}
\end{equation}

\subsubsection{Discussion}

In agreement with results shown in Fig.~\ref{Exact} (a), the expansions~\eqref{eq:COPsigmas} and \eqref{app-smallcop} clearly show that for small values of $\sigma$ and/or $\varepsilon_{\rm C}$ the COP~\eqref{eq:COPsigmas} exhibits a sharp non-linear increase when the power is decreased from its maximum value. Equation~\eqref{app-largecop}, on the other hand shows that, for moderate and large values of $\sigma$, this increase is linear with the slope determined by inverse irreversibility ratio, which is again seen in Fig.~\ref{Exact} (a). Noteworthy, all the above approximate results give correct maximum COP $\varepsilon_{\rm C}$ for vanishing cooling power $\delta P = -1$.

Let us now discuss the range of validity of the above approximations more quantitatively. To this end, we define the function
\begin{equation}
E=\frac{1}{\varepsilon_{\rm C}}\int_{-1}^0 d(\delta P) \left|\varepsilon^{\rm opt}(\delta P) - \varepsilon_{\rm approx}^{\rm opt}(\delta P)\right|,
\label{eq:Error}
\end{equation}
which measures the area in $\varepsilon^{\rm opt}$--$\delta P$ plot between the true maximum COP and its individual approximations $\varepsilon^{\rm opt}_{\rm approx}$ given by 
Eqs.~\eqref{eq:COPsigmas}, \eqref{app-smallcop}  and \eqref{app-largecop}. 

In Fig.~\ref{Test}, we show only performance of the approximations \eqref{eq:COPsigmas} and \eqref{app-largecop}. Equation~\eqref{app-smallcop} performs slightly worse than Eq.~\eqref{eq:COPsigmas} for large $\varepsilon_{\rm C}$, but it shares the same qualitative behavior. In agreement with our vague discussion above, Fig.~\ref{Test} proves that the approximation \eqref{eq:COPsigmas} works well for small $\tilde{\sigma} \gg 1$ (parameters {\bf c} and {\bf d}), but that it is also reasonable for small $\varepsilon_{\rm C}$ and large $\sigma$, yielding $\tilde{\sigma}$ of order $1$  ({\bf a}). For large values of the irreversibility ratio, Eq.~\eqref{eq:COPsigmas} yields negative values for (almost) all $\delta P$ and thus the approximation completely fails ({\bf b}). The approximation~\eqref{app-largecop}, on the other hand, performs almost perfectly for moderate and large values of $\tilde{\sigma}$  ({\bf a} and {\bf b}), but gives reasonable results also for small irreversibility ratios ({\bf c} and {\bf d}).    

\section{Conclusion and outlook}
\label{sec:conclusion}

We have derived an exact but complicated formula for maximum COP at arbitrary power for Carnot-type low-dissipation refrigerators and also three simple approximations valid for a large part of the parameter space of the model. Based on these results, we have shown that the infinitely fast non-linear increase in COP with decreasing power from its maximum value $P^\star$, routinely seen in heat engines~\cite{Whitney2014,Whitney2015,Ryabov2016,holubec2016maximum,Holubec2017,Dechant2017}, occurs in LD refrigerators only for small values of irreversibility ratio~\eqref{eq:dissipaton_ratio} or large temperature differences (which, however, lead to small ultimate upper bounds on COP $\varepsilon_{\rm C}$). For large irreversibility ratios, such an increase occurs only for small values of power, where the COP rapidly growths towards its maximum $\varepsilon_{\rm C}$.


\textcolor{black}{Our formulas for efficiency are functions of power measured in units of maximum power, which thus can further be optimized without affecting the efficiency corresponding to the fixed ratio $P/P^\star$. For slowly driven systems, one can straightforwardly use the results obtained in Ref.~\cite{Abiuso2019} for LD heat engines. For arbitrary cycle duration and a given change in the system volume (measured by the increase $\Delta S$ in system entropy during the hot isotherm), larger maximum cooling power~\eqref{final-power} corresponds to small values of reversibility parameter during the cold isotherm, $\sigma_{\rm c}$. To conclude, an ideal LD refrigerator should be based on a working fluid with small $\sigma_{\rm c}$ (yielding large maximum cooling power) and even much smaller $\sigma_{\rm h}$ (allowing one to profit from the large gain in COP while sacrificing only a small part of the maximum power). }

Our present contribution into the collection of maximum efficiencies at given power for various systems might be of immediate practical interest. \textcolor{black}{Even though the used assumptions are valid only for systems under perfect experimental control such as Brownian heat engines~\cite{schmiedl2007efficiency, holubec2015efficiency,Blickle2012,mart15}, taking into account additional sources of dissipation just leads to a decrease in efficiency. And thus the derived maximum efficiencies can be thought of as upper bounds on efficiencies even for relativistic settings. Furthermore}, our results for refrigerators could be combined with known results for heat engines to yield maximum efficiency at fixed power for absorption refrigerators, which were studied numerically in Ref.~\cite{Guo2019}. What remains to complete the collection for LD models is a derivation of maximum efficiency at fixed power for heat pumps. Both these tasks are subjects of our present research. Furthermore, it would be interesting to investigate maximum efficiency at fixed power for LD systems with respect to their dynamical stability~\cite{Gonzalez-Ayala2018,Gonzalez-Ayala2019,gonzalez2020energetic}.

\begin{acknowledgments}
VH gratefully acknowledges support by the Humboldt foundation and by the Czech Science Foundation (project No. 20-02955J). ZY is supported by the China Scholarship Council (CSC) under Grant No. 201906310136.
\end{acknowledgments}

\bibliography{References}

\appendix

\section{Minimally nonlinear irreversible model}
\label{appx:MIM}

In this appendix, we review in detail the mapping between the LD model and the minimally nonlinear irreversible (MNI) model \cite{Izu-EPL-refri, Izu-EPL-heat, Izu2015}. We proceed in two steps. First, we map the average total entropy production rate
\begin{equation}
\frac{\Delta S_{\rm tot}}{t_{\rm p}} = -\frac{Q_{\rm c}}{t_{\rm p}T_{\rm c}}+\frac{Q_{\rm h}}{t_{\rm p} T_{\rm h}}=\frac{1}{t_{\rm p}}\frac{W}{T_{\rm h}} + P \left(\frac{1}{T_{\rm h}}-\frac{1}{T_{\rm c}}\right)\label{total-entropy}
\end{equation}
for cyclic Carnot type refrigerators, depicted in Fig.~\ref{fig:T-S}, to the entropy production rate $\dot{\sigma}=J_1X_1+J_2X_2$, written as a linear combination of (generalized) fluxes $J_i$ and forces $X_i$, $i = 1,2$, used in linear irreversible thermodynamics~\cite{VandenBroeck2005,Ryabov2016}. While there is a variety of possible choices, we employ the commonly used mapping~\cite{zhang2018coefficient, izumida2014work, Izu-EPL-refri} $J_1=1/t_{\rm p}$, $X_1=W/T_{\rm h}$ and $J_2=P$, $X_2=1/T_{\rm h}-1/T_{\rm c}$. Consequently, the heat flux to the hot reservoir reads $Q_{\rm h}/t_{\rm p}=J_2+J_1X_1T_{\rm h}\equiv J_3$.

The MNI model assumes that the linear flux-force relation applied in linear irreversible thermodynamics is generalized as~\cite{Izu2015, Izu-EPL-heat, Izu-EPL-refri},
\begin{eqnarray}
{J_1} &=& {L_{11}}{X_1} + {L_{12}}{X_2},
\label{J1-Onsager}\\
{J_2} &=& {L_{21}}{X_1} + {L_{22}}{X_2} - {\gamma _{\rm c}}J_1^2.
\end{eqnarray}
Here, $L_{ij}$ $i,j = 1,2$ denote Onsager coefficients and the new term $-\gamma_{\rm c}J_1^2$, with $\gamma_{\rm c}\geq 0$, stands for a fraction of input power leaking into the cold bath. Physically, it describes frictional looses in mechanical machines or looses due to a finite resistivity in thermoelectric devices \cite{apertet2013efficiency}.

Using Eq.~(\ref{J1-Onsager}), the heat fluxes from the cold bath ($J_2$) and to the hot bath ($J_3$) read
\begin{eqnarray}
{J_2} &=& \frac{L_{21}}{L_{11}}J_1+L_{22}\left(1-q^2\right)X_2- {\gamma _{\rm c}}J_1^2,
\label{eq:J2}\\
{J_3} &=& \frac{L_{21}}{L_{11}}\frac{T_{\rm h}}{T_{\rm c}}J_1+L_{22}\left(1-q^2\right)X_2+ {\gamma _{\rm h}}J_1^2,
\label{eq:J3}
\end{eqnarray}
where $\gamma_{\rm h}J_1^2$ denotes the fraction of input power leaking into the hot reservoir. The Onsager reciprocity relations imply that the coupling strength parameter $q=L_{12}/\sqrt{L_{11}L_{22}}$ is bounded as $\left(|q|\leq 1\right)$. As the second step in the mapping, we compare Eqs.~\eqref{qh} and \eqref{qc} and \eqref{eq:J2} and \eqref{eq:J3} and try to find a mapping between the parameters. This can be done under the tight coupling condition $|q|=1$, when the flux $J_1$ and the heat fluxes $J_i$, $i=2,3$ are proportional in the linear irreversible model and efficiencies of machines based on the MNI model are largest. The result is~\cite{Izu-EPL-refri}
\begin{eqnarray}
\left( {\begin{array}{*{20}{c}}
{{L_{11}}}&{{L_{12}}}\\
{{L_{21}}}&{{L_{22}}}
\end{array}} \right) &=& \left( {\begin{array}{*{20}{c}}
{\frac{{{T_{\rm h}}}}{\lambda }}&{\frac{{{T_{\rm h}}{T_{ \rm c}}\Delta S}}{\lambda }}\\
{\frac{{{T_{\rm h}}{T_{\rm c}}\Delta S}}{\lambda }}&{\frac{{{T_{\rm h}}{{\left( {{T_{\rm c}}\Delta S} \right)}^2}}}{\lambda }}
\end{array}} \right),
\label{Onsage coeffi}\\
{\gamma _{\rm h}} &=& \frac{{{\sigma _{\rm h}}}}{\alpha},
\label{gamma-h}\\
{\gamma _{\rm c}} &=& \frac{{{\sigma _{\rm c}}}}{{1-\alpha}},
\label{gamma-c}
\end{eqnarray}
where $\gamma\equiv\gamma_{\rm h}/\gamma_{\rm c}$ and $\lambda  \equiv \sigma _{\rm h}/\alpha + \sigma _{\rm c}/(1-\alpha)$. Let us now study COP at maximum power of refrigerators based on the MNI model in terms of this mapping.

Assuming that we control either the flux $J_1$ or the corresponding thermodynamic force $X_1$, the maximum cooling power ensue from the formula $\partial J_2/\partial J_1=0$ (or, equivalently, $\partial J_2/\partial X_1=0$). We obtain the following values of model parameters at maximum cooling power~\cite{Long2018}
\begin{eqnarray}
\frac{1}{J^*_{1}} &=& \frac{2\gamma_{\rm c} L_{11}}{L_{21}},
\label{J1-max-P}\\
J^*_2 &=& \frac{L_{21}^2}{4\gamma_{\rm c} L_{11}^2},\\
\frac{J^*_2}{J^*_3 - J^*_2 } &=& \frac{\varepsilon_{\rm C}}{2+\left(1+\gamma\right)\varepsilon_{\rm C}},
\label{Onsager-COP-max-P}
\end{eqnarray}
where the last expression describes the COP at maximum power. Substituting the mapping~\eqref{Onsage coeffi}--\eqref{gamma-c} into \eqref{J1-max-P}--\eqref{Onsager-COP-max-P}, we reproduce Eqs.~(\ref{taup*})--(\ref{etatp}) corresponding to power in the LD model optimised only with respect to the duration of the isothermal branches, $t_{\rm i}$. These expressions thus still depend on the distribution of $t_{\rm i}$ between the two isotherms, $\alpha$~\cite{Izu-EPL-heat, See-AppendixC}. In order to get the final results (\ref{final-taup})--(\ref{final-COP+}) for COP at maximum power in the LD refrigerator, one thus just needs to further optimise the power~\eqref{J1-max-P} with respect to $\alpha$. This also hold for maximum COP at fixed cooling power and all other figures of merit. Indeed, substituting the mapping \eqref{gamma-h}--\eqref{gamma-c} into Eq.~(17) in Ref.~\cite{Long2018} for maximum COP at given power for MNI refrigerators and optimising the resulting expression with respect to $\alpha$, one obtains our results for the maximum COP at fixed power for LD refrigerators, described in Sec.~\ref{sec:MAXCOP}.

To conclude, LD models can exactly be mapped to MNI models with tight coupling if the later are further optimised with respect to the additional parameter $\alpha$. However, to the best of our knowledge, this possibility is usually overlooked~\cite{Long2018, Izu2015, Izu-EPL-heat, Izu-EPL-refri}. One exception where both models always give the same results are bounds on performance, obtained by taking the limits $\sigma \to 0$ and $\infty$ (or, equivalently, $\gamma \to 0$ and $\infty$). The reason is that the dependence of the mapping on $\alpha$ is lost during the limiting process.

\end{document}